\documentclass[12pt]{article}
\usepackage{amsmath}
\usepackage{amssymb,amsthm}
\usepackage[english]{babel}
\usepackage[textwidth=18.5cm,textheight=22cm]{geometry}

\newcommand{\R}{\mathbb{R}}
\newcommand{\T}{\mathbb{T}}
\newcommand{\U}{\mathbb{U}}

\newcommand{\Z}{\mathbb{Z}}

\newcommand{\mL}{\mathcal{L}}
\newcommand{\mP}{\mathcal{P}}
\newcommand{\mQ}{\mathcal{Q}}
\newcommand{\mR}{\mathcal{R}}

\newcommand{\mA}{\mathcal{A}}

\newcommand{\mM}{\mathcal{M}}

\newcommand{\ct}{\boldsymbol{\mathrm{t}}}
\newcommand{\dt}{\bar{\boldsymbol{\mathrm{t}}}}
\newcommand{\cc}{\boldsymbol{\mathrm{c}}}
\newcommand{\bc}{\boldsymbol{c}}
\newcommand{\bt}{\boldsymbol{t}}

\newcommand{\be}{\begin{equation}}
\newcommand{\ee}{\end{equation}}

\newtheorem{teh}{Theorem}

\begin{document}

\title{Semiclassical expansions  in the Toda hierarchy\\ and the hermitian matrix model
\thanks{Partially supported by  MEC
project FIS2005-00319 and ESF programme MISGAM}}
\author{L. Mart\'{\i}nez Alonso$^{1}$ and E. Medina$^{2}$
\\
\emph{$^1$ Departamento de F\'{\i}sica Te\'{o}rica II, Universidad
Complutense}\\ \emph{E28040 Madrid, Spain}\\
\emph{$^2$ Departamento de Matem\'aticas, Universidad de C\'adiz}\\
\emph{E11510 Puerto Real, C\'adiz, Spain} }
\date{} \maketitle
\maketitle \abstract{ An iterative algorithm for determining a class
of solutions of the dispersionful 2-Toda hierarchy characterized by
string equations is developed. This class includes the solution
which underlies the large $N$-limit of the Hermitian matrix model in
the one-cut case. It is also shown how the double scaling limit can
be naturally formulated in this scheme}

\vspace*{.5cm}

\begin{center}\begin{minipage}{12cm}
\end{minipage}
\end{center}
\newpage
\section{Introduction}

Since the pioneering works \cite{rus}-\cite{mar} the Toda hierarchy
has become one of the paradigmatic examples of the relevance of
integrable systems in the theory of random matrix models. As a
consequence of the activity in this field  a rich theory of the
different facets of the Toda hierarchy has been developed.

The present work is motivated by the applications of the Toda
hierarchy theory to the Hermitian matrix model. In this
model the first integrable structure which emerges is the 1-Toda hierarchy \cite{avm}
\[
\dfrac{\partial L}{\partial t_j}=[(L^j)_+,L]
\]
on semi-infinite tridiagonal matrices
\[
L=\Lambda+u_n+v_n\,\Lambda^{T},\quad n\geq 0.
\]
Here $\Lambda$ is the standard shift matrix and $\mA_+$ denotes the
upper part (above the main diagonal) of semi-infinite matrices
$\mA$. This relationship may be conveniently described by
considering infinite-dimensional deformations  of monic orthogonal
polynomials on the real line
\[
P_n(z,\bt)=z^n+\cdots,\quad \bt:=(t_1,t_2,\ldots),\quad n\geq 0,
\]
with respect to an exponential
weight:
\[
\int_{-\infty}^{\infty}P_n(z,\bt)\,P_m(z,\bt)\,e^{V(z,\bt)}\,
dz=h_n(\ct)\delta_{nm},\quad V(z,\bt):=\sum_{k\geq 1}(t_k+c_k)\,z^k,
\]
where $\bc:=(c_1,c_2,\ldots)$ is a given set of complex constants.
It turns out  (see for instance \cite{avm}) that the functions
\[
\psi_n(z,\bt):=P_n(z,\bt)\,\exp(\sum_{k\geq 1}z^k\,t_k),\quad n\geq
0,
\]
satisfy the linear system of the semi-infinite 1-Toda hierarchy
\[
\dfrac{\partial \psi_n}{\partial t_k}=(L^k)_+\,\psi_n,\quad L\,\psi_n=z\,\psi_n,\quad k\geq 1,\quad n\geq 0,
\]
and have  a $\tau$-function representation
\[
\psi_n(z,\bt)=\dfrac{\tau_n(\bt-[z^{-1}])}{\tau_n(\bt)}\,z^n\,\exp(\sum_{k\geq 1}z^k\,t_k),
\]
provided by the partition function of the Hermitian matrix model
\begin{equation}\label{par}
\tau_N(\bt)=Z_N(\bt):=\int_{\mathbb{R}^N}\prod_{k=1}^{N}\Big(d\,x_k\,e^{V(x_k,\bt)}\Big)(\Delta(x_1,\cdots,x_N))^2,
\end{equation}
where $\Delta(x_1,\cdots,x_n):=\prod_{i>j}(x_i-x_j)$.

\vspace{0.3cm}

Many exciting properties of the Hermitian  model emerge in the
analysis of its large $N$-limit \cite{fok}-\cite{dei}. One of the
main tools supplied by the theory of the Toda hierarchy for such
analysis \cite{rus} is the use of a pair of  constraints called
\emph{string equations}
\begin{equation}\label{diss}
L=\Lambda+u_n+v_n\,\Lambda^{T},\quad M=\sum_{k\geq
1}k\,(t_k+c_k)(L^{k-1})_+
\end{equation}
which are satisfied by the canonically conjugated operators
\[
L\,\psi_{n}=z\,\psi_{n},\quad
\dfrac{\partial\psi_n}{\partial z}=M\,\psi_n.
\]
The present paper deals with the analysis of the  large $N$-limit
of the partition function
\begin{equation}\label{lnl}
Z_N(N\,\ct)
=\int_{\mathbb{R}^N}\prod_{k=1}^{N}\Big(d\,x_k\,e^{N\,V(x_k,\ct)})\Big)(\Delta(x_1,\cdots,x_N))^2.
\end{equation}
Here a small parameter $\epsilon:=1/N$ and rescaled variables $
\ct:=\epsilon\,\bt$ and   constants $\cc:=\epsilon\,\bc$ have been
introduced. Since $\epsilon$  plays the role of the Planck constant
$\hbar$, expansions in powers of $\epsilon$ are referred to as
\emph{semiclassical} expansions. The same slow variables
$\ct=\epsilon\,\bt$ together with a continuous variable
$x:=\epsilon\,n$ are introduced to pass from the Toda hierarchy to
its  \emph{dispersionful} formulation \cite{tt4}, which provides an
interpolated continuous version of the Toda hierarchy. In this way,
and due to the fact that  $\tau_n(\bt)=Z_n(\bt)$ is a
$\tau$-function of the semi-infinite 1-Toda hierachy, it is natural
to expect that a $\tau$-function $\tau(\epsilon,x,\ct)$  of the
dispersionful 1-Toda hierarchy verifying
\begin{equation}\label{rel}
\tau(\epsilon,\epsilon\,n,\ct)=Z_n(N\,\ct),
\end{equation}
should describe the large $N$-limit of the Hermitian model. In this paper we are concerned with the characterization of this solution of the dispersionful 1-Toda hierarchy.

The main result of our work
 is a scheme for obtaining solutions of the dispersionful 2-Toda hierarchy satisfying the system of string equations
\begin{equation}\label{sei}
\mL=\bar{\mL},\quad \mM+
F(\mL)=\overline{\mM}+\overline{F}(\bar{\mL}),
\end{equation}
where $(\mL,\mM)$ and $(\bar{\mL},\overline{\mM})$ denote  two
pairs of Lax-Orlov operators  and $(F,\,\bar{F})$ are two arbitrary functions.  The first string equation represents
the 1-Toda reduction condition and is satisfied by Lax operators of
the form
\[
\mL=\bar{\mL}=\Lambda+u+v\,\Lambda^{-1},
\]
where now $\Lambda:=\exp{(\epsilon\,\partial_x)}$, and $(u,v)$ are
characterized by semiclassical expansions
\begin{equation}\label{semc}
u=\sum_{k\geq 0}\epsilon^k\,u^{(k)},\quad v=\sum_{k\geq
0}\epsilon^{2k}\,v^{(2k)}.
\end{equation}
The point is that for $x=1$ and $\bar{F}\equiv 0$ the constraints
\eqref{sei} interpolate \eqref{diss}, so that the corresponding
solution of the dispersionful 1-Toda hierarchy is a candidate to
the solution underlying the large $N$-limit of the Hermitian model.
That it is the only possible candidate can be argued as follows:
\begin{enumerate}
\item Recent research \cite{dub1}-\cite{dub3} proved that the solutions
of an extended version of the dispersionful 1-Toda hierarchy  are
determined by the leading order terms $(u^{(0)}, v^{(0)})$. In fact,
the coefficients $(u^{(k)}, v^{(2k)})$  are rational functions of
$(u^{(0)}, v^{(0)})$ and their $x$-derivatives (quasi-triviality
property).
\item As it is shown in this paper, the terms $(u^{(0)}, v^{(0)})$
of the solution of \eqref{sei} coincide with those characterizing
the leading order in the large $N$-expansion (planar limit) of the
Hermitian model.
\end{enumerate}

  Our strategy is inspired by previous results \cite{mel3}-\cite{mano2} on solution
methods for dispersionless string equations. We also develop some
useful standard technology of the theory of Lax equations
\cite{che}-\cite{kup} in the context of the dispersionful 1-Toda
hierarchy. Thus we introduce two generating functions $\R$ and $\T$
related to the resolvent of the Lax operator which play a crucial role in our analysis.

The paper is organized as follows:

In the next section  the basic theory of the dispersionful 2-Toda
hierarchy and the method of  string equations are discussed. In
Section 3 we deal with  the dispersionful 1-Toda hierarchy and its
relationship with the Hermitian matrix model from the point of view
of the continuous string equations \eqref{sei}. The generating
functions $\R$ and $\T$ are introduced and are characterized by two
important identities. Our main results are derived in Section 4
where a scheme for solving the string equations \eqref{sei} in terms
of semiclassical expansions  is provided. In particular we prove
that the leading terms of these expansions characterize the planar
limit of the Hermitian matrix model.  In Section 5 it is showed how
the double scaling limit method can be naturally implemented in our
scheme.

Applications of our method to normal matrix models which are also
related to string equations of the Toda hierarchy
\cite{zab1}-\cite{kaz} will be considered elsewhere.

\section{String equations in the dispersionful 2-Toda hierarchy}

\subsection{The dispersionful 2-Toda hierarchy}

 The formulation of the dispersionful
2-Toda  hierarchy \cite{tt4} uses operators of the form  \begin{equation}\label{opint}
\mathcal{A}=\sum_{j\in\Z}\,a_j(\epsilon,x,\ct,\dt)\,\Lambda^j,\quad
\Lambda:=\exp{(\epsilon\,\partial_x)},
\end{equation}
where $x$ is a complex variable and  the coefficients are in turn  series  in  the  small parameter $\epsilon$
\[
a_j(\epsilon,x,\ct,\dt)=\sum_{k\in\Z}\epsilon^k\,a_j^{(k)}(x,\ct,\dt).
\]
Here  $\ct=(t_1,t_2,\ldots)$ and
$\dt=(\bar{t}_1,\bar{t}_2,\ldots)$ denote two infinite sets of
complex variables. The order  in $\epsilon$ of $\mathcal{A}$ is
defined by
\[
\mbox{ord}_{\epsilon}(\mathcal{A}):=\mbox{max$\{-k\,|\,a_j^{(k)}(x,\ct,\dt)\neq 0\}$}.
\]
For example $\mbox{ord}_{\epsilon}(\epsilon)=-1$ and $\mbox{ord}_{\epsilon}(\Lambda)=0$. In particular, zero-order operators are those with regular coefficents $a_j$ as $\epsilon\rightarrow 0$. As usual the $\mathcal{A}_{\pm}$ parts of $\mathcal{A}$ will denote  the
truncations of $\Lambda$-series in the positive and strictly
negative power terms, respectively.  Given a function $w$ depending on $x$, the following notation convention will be henceforth used
\[
w_{[r]}:=\Lambda^r w=w(x+r\,\epsilon),\quad r\in\Z.
\]

 The dispersionful 2-Toda hierarchy can be formulated
in terms of
a pair of formal wave functions of the form
\begin{align}\label{quas}
\nonumber &\Psi=\exp{\Big(\dfrac{1}{\epsilon}\,\mathbb{S}\Big)},\quad
\mathbb{S}
=\sum_{j=1}^\infty
t_j z^{j}+x\,\log{z}-\sum_{j\geq 1}\dfrac{1}{j\,z^j}S_{j+1},
\\\\\
\nonumber &\bar{\Psi}=z^{-1}\,\exp{\Big(\dfrac{1}{\epsilon}\,\bar{\mathbb{S}}\Big)},\quad
\bar{\mathbb{S}}
=\sum_{j=1}^\infty
\bar{t}_j z^{j}-x\,\log{z}-\bar{S}_0-\sum_{j\geq 1}\dfrac{1}{j\,z^j}\bar{S}_{j+1},
\end{align}
where
\[
S_j=\sum_{k\geq 0}\epsilon^k\,S_j^{(k)}(x,\ct,\dt),\quad
\bar{S}_j=\sum_{k\geq 0}\epsilon^k\,\bar{S}_j^{(k)}(x,\ct,\dt),
\]
These functions $\mathbf{\Psi}=\Psi,\,\overline{\Psi}$ are assumed to satisfy  the linear system
\begin{equation}\label{contwave1i}
\epsilon\,\dfrac{\partial \mathbf{\Psi}}{\partial
t_j}=(\mL^j)_+\,\mathbf{\Psi}, \quad \epsilon\,\dfrac{\partial
\mathbf{\Psi}}{\partial \bar{t}_j}=(\bar{\mL}^j)_-\,\mathbf{\Psi},
\end{equation}
where the Lax operators $\mL$ and $\bar{\mL}$ are  determined by the
equations
\begin{equation}\label{zopi}
\mL\,\Psi=z\,\Psi,\quad
\bar{\mL}\,\overline{\Psi}=z\,\overline{\Psi},
\end{equation}
and are assumed \cite{tt4}  to be of  zero order in $\epsilon$. We will also use the Orlov operators $\mM$ and $\overline{\mM}$
characterized by
\begin{equation}\label{zopi1}
\mM\,\Psi=\epsilon\,\dfrac{\partial \Psi}
{\partial z};\quad
\overline{\mM}\,\overline{\Psi}=\epsilon\,\dfrac{\partial \overline{\Psi}}
{\partial z},
\end{equation}
which satisfy
\[
[\mL,\mM]=[\overline{\mL},\overline{\mM}]=\epsilon.
\]

Using \eqref{quas} and \eqref{zopi}-\eqref{zopi1} one sees that the following expansions follow
\begin{align}\label{sati3}
\nonumber &\mL=\Lambda+u_0+u_1\,\Lambda^{-1}+\cdots,\quad
\mM=\sum_{j=1}^\infty
j\,t_j \mL^{j-1}+x\,\mL^{-1}+\sum_{j\geq 1}S_{j+1}\,\mL^{-j-1}\\\\
\nonumber
&\bar{\mL}=\bar{u}_{-1}\,\Lambda^{-1}+\bar{u}_0+\bar{u}_1\,\Lambda+\cdots,\quad
\overline{\mM}=\sum_{j=1}^\infty j\,\bar{t}_j
\bar{\mL}^{j-1}-(x+\epsilon)\bar{\mL}^{-1}+\sum_{j\geq
1}\bar{S}_{j+1}\,\bar{\mL}^{-j-1}.
\end{align}
Furthermore, \eqref{contwave1i} can be rewritten in Lax form as
\begin{equation}\label{laxxi}
\epsilon\,\dfrac{\partial K}{\partial t_j}=[(\mL^j)_+,K],\quad
\epsilon\,\dfrac{\partial K}{\partial
\bar{t}_j}=[(\bar{\mL}^j)_-,K],
\end{equation}
where $K=\mL,\,\mM,\,\bar{\mL},\,\overline{\mM}$.

\vspace{0.2cm}

There is also a $\tau$-function representation of the wave functions \cite{tt4}
\begin{align}\label{tau}
\nonumber
&\Psi=\exp{\Big(\dfrac{1}{\epsilon}\,(\sum_{j=1}^\infty
t_j z^{j}+x\,\log{z})\Big)}\,
\dfrac{\tau(\epsilon,x,\ct-\epsilon\,[z^{-1}],\dt)}{\tau(\epsilon,x,\dt)},
\\\\\
\nonumber
&\bar{\Psi}=z^{-1}\,\exp{\Big(\dfrac{1}{\epsilon}\,(\sum_{j=1}^\infty
\bar{t}_j z^{j}-x\,\log{z})\Big)}\,\dfrac{\tau(\epsilon,x+\epsilon,\ct,\dt-\epsilon\,[z^{-1}])}{\tau(\epsilon,x,\ct,\dt)},
\end{align}
where $[z^{-1}]:=(1/z,\,1/2z^2,\,1/3z^3,\ldots)$ and $\tau$ is of the form.
\begin{equation}\label{tau0}
\tau=\exp{\Big(\dfrac{1}{\epsilon^2}\,\mathbb{F}\Big)},\quad
\mathbb{F}=\sum_{k\geq 0}\epsilon^k\,F^{(k)}(x,\ct,\dt).
\end{equation}

The dispersionful 2-Toda hierachy arises as a continuum limit of
the standard 2-Toda hierachy \cite{ueta} in which  the
standard discrete variable $n$ is substituted by a  continuous variable $x$ and
two sets of  \emph{fast} continuous variables
$\bt:=\epsilon^{-1}\ct,\,\bar{\bt}:=\epsilon^{-1}\dt$ are introduced.   Thus, the main dynamical objects ($\tau$-functions, wave
functions, Lax and Orlov operators) of both hierarchies are
related by
\begin{equation}\label{rel0}
\tau(\epsilon,\epsilon\,n,\epsilon\,\bt,\epsilon\,
\bar{\bt})=\tau_n(\bt,\bar{\bt})
\end{equation}
and
\begin{align}\label{rel1}
\nonumber& \Psi(z,\epsilon,\epsilon\,n,\epsilon\,\bt,\epsilon\,
\bar{\bt})=\psi_n(z,\bt,\bar{\bt}),\quad \quad \,\,\,\bar{\Psi}(z,\epsilon,\epsilon\,n,\epsilon\,\bt,\epsilon\,
\bar{\bt})=\bar{\psi}_n(z,\bt,\bar{\bt}),\\
& \mL(z,\epsilon,\epsilon\,n,\epsilon\,\bt,\epsilon\,
\bar{\bt})=L(z,n,\bt,\bar{\bt}) ,\quad \quad\,\,\,
\bar{\mL}(z,\epsilon\,n,\epsilon\,\bt,\epsilon\,
\bar{\bt})=\bar{L}(z,n,\bt,\bar{\bt}),\\
\nonumber & \mM(z,\epsilon,\epsilon\,n,\epsilon\,\bt,\epsilon\,
\bar{\bt})=\epsilon\, M(z,n,\bt,\bar{\bt}),\quad
\overline{\mM}(z,\epsilon,\epsilon\,n,\epsilon\,\bt,\epsilon\,
\bar{\bt})=\epsilon\,\overline{M}(z,n,\bt,\bar{\bt}).
\end{align}

Our subsequent analysis uses an important result proved by
Takasaki and Takebe (Proposition 2.7.11. in  \cite{tt4})

\begin{teh} Suppose that
\begin{align*}
&\mP(\epsilon,x\,\Lambda^{-1},\Lambda)=\sum_{k\in\Z}p_k(\epsilon,x\,\Lambda^{-1})\,\Lambda^k,\quad
\mQ(\epsilon,x\,\Lambda^{-1},\Lambda)=\sum_{k\in\Z}q_k(\epsilon,x\,\Lambda^{-1})\,\Lambda^k,\\\\
&\overline{\mP}(\epsilon,x\,\Lambda^{-1},\Lambda)=\sum_{k\in\Z}\bar{p}_k(\epsilon,x\,\Lambda^{-1})\,\Lambda^k,\quad
\overline{\mQ}(\epsilon,x\,\Lambda^{-1},\Lambda)=\sum_{k\in\Z}\bar{q}_k(\epsilon,x\,\Lambda^{-1})\,\Lambda^k,
\end{align*}
are operators of zero order in $\epsilon$
verifying
\[
[\mP,\mQ]=[\overline{\mP},\overline{\mQ}]=\epsilon
\]
If $(\mL,\mM)$ and $(\bar{\mL},\overline{\mM})$ are operators of zero order in $\epsilon$ of the form \eqref{sati3} which satisfy the pair of constraints
\begin{equation}\label{string}
\mP(\epsilon,\mM,\mL)=\overline{\mP}(\epsilon,\overline{\mM},
\bar{\mL}),\quad
\mQ(\epsilon,\mM,\mL)=\overline{\mQ}(\epsilon,\overline{\mM},\bar{\mL}),
\end{equation}
and
\begin{equation}\label{can}
[\mL,\mM]=[\overline{\mL},\overline{\mM}]=\epsilon,
\end{equation}
then $(\mL,\mM)$ and $(\bar{\mL},\overline{\mM})$ are solutions of the Lax equations \eqref{laxxi} of the 2-Toda hierarchy.

\end{teh}

Constraints of the form \eqref{string} are called \emph{string equations}. In this work we are interested in the particular  example given by
\begin{equation}\label{tridiagonal}
\begin{cases}
\mL=\bar{\mL},\\
\mM+ F(\mL)=\overline{\mM}+\overline{F}(\bar{\mL}),
\end{cases}
\end{equation}
where $F(\mL)$ and $\overline{F}(\bar{\mL})$ are arbitrary functions of the form
\[
F(\mL):=\sum_{j\geq 1}j\,c_j\,\mL^{j-1},\quad \bar{F}(\bar{\mL}):=\sum_{j\geq 1}j\,\bar{c}_j\,\bar{\mL}^{j-1}.
\]

\section{The dispersionful 1-Toda hierarchy and the Hermitian model}

The first string equation in \eqref{tridiagonal} represents the so called \emph{tridiagonal} (1-Toda) reduction of the dispersionful 2-Toda hierarchy and implies the following form of the Lax operators
\begin{equation}\label{tri}
\mL=\bar{\mL}=\Lambda+u+v\,\Lambda^{-1}.
\end{equation}
Thus, as a consequence of the Lax equations, $u$ and $v$ depend on
$(\ct,\dt)$ through the combination $\ct-\dt$.
 Moreover
\eqref{tri} implies
\begin{equation}\label{tri1}
(\Lambda+u+v\,\Lambda^{-1})\,\Psi=z\,\Psi,\quad
(\Lambda+u+v\,\Lambda^{-1})\overline{\Psi}=z\,\overline{\Psi},
\end{equation}
so that
\begin{equation}\label{d6q}
u=\epsilon^{-1}\,(S_{2[1]}-S_2),\quad
\log{v}=\epsilon^{-1}\,(\bar{S}_{0[-1]}-\bar{S}_0).
\end{equation}

In order to solve the string equations \eqref{tridiagonal} it is
required to characterize the action of the operators $(\mL^j)_+$ and
$(\bar{\mL}^j)_-$ on the wave functions $\Psi$ and
$\overline{\Psi}$. This calculation is also needed to determine the
integrable systems of the dispersionful 1-Toda hierarchy. We start
by introducing the two series in $z$
\begin{equation}\label{ps}
p(z)=z-u+\mathcal{O}\Big(\dfrac{1}{z}\Big),\quad
\bar{p}(z)=\dfrac{v_{[1]}}{z}+\mathcal{O}\Big(\dfrac{1}{z^2}\Big),\quad
z\rightarrow\infty,
\end{equation}
satisfying
\begin{equation}\label{tri2}
\Lambda\,\Psi=p(z) \,\Psi,\quad
\Lambda\,\overline{\Psi}=\bar{p}(z) \,\overline{\Psi},
\end{equation}
which according to \eqref{tri1} are determined by
\begin{equation}\label{tri3}
\boldsymbol{p}(z)+u+\dfrac{v}{\boldsymbol{p}_{[-1]}(z)}=z,
\end{equation}
where $\boldsymbol{p}=p,\,\bar{p}$. By
using \eqref{tri1} it is clear that there are functions
$\alpha_j,\beta_j,\bar{\alpha}_j,\bar{\beta} _j$, which depend
polynomially in $z$, such that
\begin{eqnarray}\label{tri2e}
\nonumber \epsilon\,\dfrac{\partial \mathbf{\Psi}}{\partial
t_j}=(\mL^j)_+\mathbf{\Psi}=\alpha_j\,\mathbf{\Psi}+\beta_j\,\Lambda\mathbf{\Psi}
=(\alpha_j+\beta_j\,\boldsymbol{p})\mathbf{\Psi},\\\\
\nonumber\epsilon\,\dfrac{\partial \mathbf{\Psi}}{\partial
\bar{t}_j}=(\bar{\mL}^j)_-\mathbf{\Psi}=\bar{\alpha}_j\,\mathbf{\Psi}+\bar{\beta}
_j\,\Lambda \mathbf{\Psi} =(\bar{\alpha}_j+\bar{\beta}
_j\,\boldsymbol{p})\mathbf{\Psi},
\end{eqnarray}
and
\begin{equation}\label{tri5}
\bar{\alpha}_j=z^j-\alpha_j,\quad \bar{\beta}_j=-\beta_j.
\end{equation}
Hence, we have
\begin{equation}\label{tri3e}
\alpha_j+\beta_j\,p=\partial_{t_j} \mathbb{S}(z)=
z^j+\mathcal{O}\Big(\dfrac{1}{z}\Big),\quad
\alpha_j+\beta_j\,\bar{p}=\partial_{t_j}\bar{\mathbb{S}} (z)
=-\partial_{t_j} \bar{S}_0+ \mathcal{O}\Big(\dfrac{1}{z}\Big),
\end{equation}
so that
\begin{equation}\label{tri4e}
\alpha_j=\dfrac{1}{2}\Big(z^j-\partial_{t_j}
\bar{S}_0-(\beta_j\,(p+\bar{p}))_\oplus\Big),\quad
\beta_j=\Big(\dfrac{z^j}{p-\bar{p}}\Big)_\oplus,
\end{equation}
where $(\;)_\oplus$ and $(\;)_\ominus$ stand for the projections of $z$-series on the subspaces generated by
the positive and strictly negative powers , respectively.

At this point it is useful to   introduce  the generating functions
\begin{equation}\label{tri5a}
\R:=\dfrac{z}{p-\bar{p}}=\sum_{k\geq
0}\dfrac{R_k(u,v)}{z^k},\quad
\T:=\dfrac{p+\bar{p}}{p-\bar{p}}=\sum_{k\geq
0}\dfrac{T_k (u,v)}{z^k},\quad R_0=T_0=1.
\end{equation}
By  substituting  $p$ and $\bar{p}$ by their expressions in terms of
$\R$ and $\T$ in the identities
\begin{equation}\label{tri4}
u=z+\dfrac{p\,p_{[-1]}-\bar{p}\,\bar{p}_{[-1]}}{\bar{p}_{[-1]}-p_{[-1]}},\quad
v=\dfrac{\bar{p}-p}{\bar{p}_{[-1]}-p_{[-1]}}\,\bar{p}_{[-1]}p_{[-1]},
\end{equation}
we obtain the following relations
\begin{equation}\label{tri7}
\begin{cases}
\T_{[1]}+\T+\dfrac{2}{z}(u_{[1]}-z)\,\R_{[1]}=0,\\\\
\T^2-\dfrac{4}{z^2}\,v_{[1]}\,\R\,\R_{[1]}=1,
\end{cases}
\end{equation}
which allow us to compute recursively the coefficients of the series \eqref{tri5a}  as polynomials in $u$, $v$
and their $x$-translations $u_{[r]}$ and $v_{[r]}$. Indeed, the
system  \eqref{tri7} implies
 \be\label{RSm}\everymath{\displaystyle}
\begin{cases}
2T_{k+1}=-\sum_{ i+j=k+1;i,j\geq 1}T_i\,T_j+4v_{[1]}\sum_{i+j=k-1}R_i\,R_{j[1]},\\  \\
R_{k+1}=u\,R_k+\frac{1}{2}[T_{k+1}+T_{k+1\,[-1]}],\quad .
\end{cases}
\ee For example, the  first few  coefficients are:
\begin{align*}
T_1&=0,\quad R_1=u,\quad T_2=2v_{[1]},\quad R_2=u^2+v_{[1]}+v,\\
T_3&=2v_{[1]}(u+u_{[1]}),\quad
R_3=u^3+2uv_{[1]}+2uv+u_{[1]}v_{[1]}+u_{[-1]}v,\\
T_4&=2v_{[1]}\,(u_{[1]}^2+uu_{[1]}+u^2+v_{[2]}+v_{[1]}+v).
\end{align*}

In this way, by taking into account the second equation of
\eqref{RSm}, one finds
\begin{align}\label{tri6}
\nonumber
\partial_{t_j}\,\mathbb{S}(z)&=\alpha_j+\beta_j\,p=z^j-\dfrac{1}{2}\partial_{t_j}
\bar{S}_0-\dfrac{z}{2\,\R}\,(z^{j-1}\,\R)_\ominus+
\Big(\dfrac{z}{2\,\R}\,\T\,(z^{j-1}\,\R)_\oplus\Big)_\ominus
\\\\
\nonumber &=z^j-\dfrac{1}{2}(\partial_{t_j}
\bar{S}_0+R_j)-\dfrac{1}{2\,z}T_{j+1[-1]}+\mathcal{O}\Big(\dfrac{1}{z^2}\Big),
\end{align}
so that
\[
\partial_{t_j}
\bar{S}_0=-R_j,\quad\partial_{t_j} S_2=\dfrac{1}{2}\,T_{j+1[-1]}
\]
and then from \eqref{d6q} we get that the flows of the dispersionful $1$-Toda hierarchy can be expressed as
\begin{equation}\label{tri8}
\epsilon\,\partial_{t_j}u=\dfrac{1}{2}\,(T_{j+1}-T_{j+1[-1]}),\quad
\epsilon\,\partial_{t_j}v=v\,(R_j-R_{j[-1]}).
\end{equation}

Furthermore,  our calculation implies the following useful relations
\begin{align}\label{4.20}
\nonumber
&(\mL^j)_-\Psi=\Big(-\dfrac{1}{2}R_j+\dfrac{z}{2\,\R}\,(z^{j-1}\,\R)_\ominus-
\Big(\dfrac{z}{2\,\R}\,\T\,(z^{j-1}\,\R)_\oplus\Big)_\ominus\Big)\,\Psi,
\\\\
\nonumber &(\mL^j)_+\overline{\Psi}=\Big(
\dfrac{1}{2}R_j+\dfrac{z}{2\,\R}\,(z^{j-1}\,\R)_\ominus+
\Big(\dfrac{z}{2\,\R}\,\T\,(z^{j-1}\,\R)_\oplus\Big)_\ominus\Big)\,\overline{\Psi},
\end{align}
for $j\geq 1$. In particular, by taking the second equation of \eqref{RSm} into account one finds that as $z\rightarrow\infty$
\begin{align}\label{4.2}
\nonumber
&(\mL^j)_-\Psi=\Big(\dfrac{1}{2z}\,T_{j+1[-1]}+\mathcal{O}\Big(\dfrac{1}{z^2}\Big)\Big)\,
\Psi,
\\\\
\nonumber
&(\mL^j)_+\overline{\Psi}=\Big(R_j+\dfrac{1}{2z}\,T_{j+1}+\mathcal{O}\Big(\dfrac{1}{z^2}\Big)\Big)
\overline{\Psi}.
\end{align}
We observe that since $R_0=1,\; T_1=0$ these last equations hold for $j\geq 0$.

By following the analysis of \cite{kup} it can be seen that $\R$ and
$\T$ are closely related to the resolvent of the Lax operator $\mL$
\[
\mR:=(z-\mL)^{-1}.
\]
Thus, from Lemmas 3.5 and 3.18 of \cite{kup} one proves that
\[
\Big(z-\dfrac{2\,\R}{(1+\,\T)}\,\Lambda\Big)\,\mR_+=\R,\quad
\mbox{Res}\, \mR_+=\dfrac{\R}{z},
\]
where $\mbox{Res}(\sum c_k\,\Lambda^k):=c_0$

\subsubsection*{$\tau$-function representation}

 It follows from \eqref{tau} and\eqref{tri1} that the functions $u$ and $v$ can be written in terms of the $\tau$-function as
\begin{equation}\label{uvt}
u=\epsilon\,\dfrac{\partial}{\partial t_1}\,\log\,
\dfrac{\tau(\epsilon,x+\epsilon,\ct)}{\tau(\epsilon,x,\ct)},\quad
v=\dfrac{\tau(\epsilon,x+\epsilon,\ct)\,\tau(\epsilon,x-\epsilon,\ct)
}{\tau^2(\epsilon,x,\ct)},
\end{equation}
where we have set $\ct-\dt\rightarrow \ct$.
 On the other hand, it can be proved
\cite{dub2}-\cite{dub3} that the $\epsilon$-expansion of the
$\tau$-functions of the dispersionful 1-Toda hierarchy is of the
form
\begin{equation}\label{tau1}
\tau=\exp{\Big(\dfrac{1}{\epsilon^2}\,\mathbb{F}\Big)},\quad
\mathbb{F}=\sum_{k\geq 0}\epsilon^{2k}\,F^{(2k)}.
\end{equation}
As a consequence $u$ and $v$  can be expanded as
\begin{equation}\label{uve}
u=\sum_{k\geq 0}\epsilon^{k}\,u^{(k)},\quad v=\sum_{k\geq
0}\epsilon^{2k}\,v^{(2k)}.
\end{equation}

\vspace{0.3cm}

Let us introduce the  \emph{reduced} $\mathbb{S}$ and $\mathbb{M}$
functions
\[
\mathbb{S}_r:=-\sum_{j\geq 1}\dfrac{1}{j\,z^j}S_{j+1},\quad
\mathbb{M}_r:=\dfrac{\partial \mathbb{S}_r}{\partial z}.
\]
From \eqref{tau} we see that
\begin{equation}\label{ree}
\mathbb{F}(\epsilon,x,\ct-\epsilon\,[z^{-1}])-\mathbb{F}(\epsilon,x,\ct)=\epsilon\,\mathbb{S}_r(\epsilon,z,x,\ct),
\end{equation}
and by differentiating this equation with respect to $z$ we obtain
\begin{equation}\label{ree1}
\sum_{j\geq 1}\dfrac{1}{z^{j+1}}\dfrac{\partial}{\partial
t_j}\,\mathbb{F}(\epsilon,x,\ct)
=\mathbb{M}_r(\epsilon,z,x,\ct+\epsilon\,[z^{-1}]).
\end{equation}
This identity can be rewritten in in a more convenient form.
Indeed \eqref{ree} implies
\[
\mathbb{S}_r(\epsilon,z,x,\ct)-\mathbb{S}_r(\epsilon,z,x,\ct-\epsilon\,[z'^{-1}])=\mathbb{S}_r(\epsilon,z',x,\ct)-\mathbb{S}_r(\epsilon,z',x,\ct-\epsilon\,[z^{-1}]),
\]
and by differentiating with respect to $z$ and then taking the limit $z'\rightarrow z$ one finds
\[
\mathbb{M}_r(\epsilon,z,x,\ct-\epsilon\,[z^{-1}])=
\mathbb{M}_r(\epsilon,z,x,\ct)+\epsilon\, \sum_{j\geq
1}\dfrac{1}{z^{j+1}}\dfrac{\partial \mathbb{S}_r}{\partial
t_j}(\epsilon,z,x,\ct-\epsilon\,[z^{-1}]).
\]
Thus \eqref{ree1} becomes
\begin{equation}\label{ree2}
\sum_{j\geq 1}\dfrac{1}{z^{j+1}}\dfrac{\partial \mathbb{F}}{\partial t_j}=\mathbb{M}_r-\epsilon\, \sum_{j\geq 1}\dfrac{1}{z^{j+1}}\dfrac{\partial \mathbb{S}_r}{\partial t_j}.
\end{equation}

\subsection{The Hermitian matrix model}

Let us write the partition function of the Hermitian matrix
model in terms of slow variables
$\ct:=\epsilon\,\bt$, where $\epsilon=1/N$
\begin{equation}\label{mat}
Z_n(N\,\ct)=\int_{\mathbb{R}^n}\prod_{k=1}^{n}\Big(d\,x_k\,e^{N\,V(x_k,\ct)})\Big)(\Delta(x_1,\cdots,x_n))^2,\quad
V(z,\ct):=\sum_{k\geq 1}(t_k+\cc_k)\,z^k.
\end{equation}
The  large $N$-limit of the model is determined by
the asymptotic expansion  of $Z_n(N\,\ct)
$ for $n=N$ as $N\rightarrow \infty$
\begin{equation}
Z_N(N\,\ct)
=\int_{\mathbb{R}^N}\prod_{k=1}^{N}\Big(d\,x_k\,e^{N\,V(x_k,\ct)})\Big)(\Delta(x_1,\cdots,x_N))^2,
\end{equation}

 It is well-known \cite{avm} that   $Z_n(\bt)$ is a $\tau$-function of the semi-infinite 1-Toda hierachy , then  in view of \eqref{rel0} we may look for a $\tau$-function $\tau(\epsilon,x,\ct)$ of the dispersionful 1-Toda hierarchy
verifying
\begin{equation}\label{rel}
\tau(\epsilon,\epsilon\,n,\ct)=Z_n(N\,\ct),
\end{equation}
and consequently
\begin{equation}\label{rel1}
\tau(\epsilon,1,\ct)=Z_N(N\,\ct).
\end{equation}
The point is that for
\begin{equation}\label{rest}
x=1,\quad \bar{t}_j=\bar{c}_j=0,\quad j\geq 1,
\end{equation}
the system  \eqref{tridiagonal} of continuous string equations
interpolates the  discrete system \eqref{diss}. Hence the solution
 of the dispersionful 1-Toda hierarchy provided by \eqref{tridiagonal} can be expected to correspond to the
$\tau$-function verifying \eqref{rel} and, as a consequence, to
describe the the large $N$-limit of the Hermitian matrix model.

We may express the  $1/N$-expansions of the main objects of the
hermitian matrix model in terms of  objects in the dispersionful
1-Toda hierarchy. For instance, from \eqref{ree2} the
\emph{one-loop correlator}
\[
W(z):=\dfrac{1}{N}\,\sum_{j\geq 0}\dfrac{1}{z^{j+1}}\langle tr
M^j\rangle=\dfrac{1}{z}+\dfrac{1}{N^2}\,\sum_{j\geq
1}\dfrac{1}{z^{j+1}}\, \dfrac{\partial \log\,Z_N(N\,\ct)}{\partial
t_j},
\]
becomes
\begin{equation}\label{ree3}
W(z)=\dfrac{1}{z}+\mathbb{M}_r(\epsilon,z,1,\ct)-\epsilon\, \sum_{j\geq 1}\dfrac{1}{z^{j+1}}\dfrac{\partial \mathbb{S}_r}{\partial t_j}(\epsilon,z,1,\ct).
\end{equation}
Loop correlators of higher order can be obtained from $W(z)$ by application of the \emph{loop-insertion} operator $d/d\,V(z)$ \cite{amb}
\begin{align*}
W(z_1,\ldots,z_s)&=\dfrac{d}{d\,V(z_s)}\cdots\dfrac{d}{d\,V(z_{2})}\,W(z_1)\\
\dfrac{d}{d\,V(z)}:&=\sum_{j\geq 1}\dfrac{1}{z^{j+1}}\dfrac{\partial}{\partial t_j}.
\end{align*}

\section{Semiclassical expansions }

We now turn to the solutions of the system of string equations
\eqref{tridiagonal}. The first  equation is solved by setting
\[
\mL=\bar{\mL}=\Lambda+u+v\,\Lambda^{-1},
\]
which is in agreement with the  asymptotic form
\eqref{sati3} required for $\mL$ and $\bar{\mL}$.

 Let us consider the second string equation of
\eqref{tridiagonal}. We look for solutions $\mM$ and
$\overline{\mM}$ verifying asymptotic expansions of the form
\eqref{sati3}. To this end we first set
\[
\mM+ F(\mL)=\overline{\mM}+\bar{F}(\overline{\mL})= \sum_{j=1}^\infty
j\,(t_j+c_j)\,(\mL^{j-1})_++ \sum_{j=1}^\infty
j\,(\bar{t}_j+\bar{c}_j)\,(\overline{\mL}^{j-1})_-,
\]
which, taking into account the first string equation, leads to
\begin{align}\label{4.3}
\nonumber &\mM=\sum_{j=1}^\infty
j\,t_j\,\mL^{j-1}+\sum_{j=1}^\infty
j\,\Big((\bar{t}_j+\bar{c}_j)-(t_j+c_j)\Big)\,(\mL^{j-1})_-,\\\\
\nonumber &\overline{\mM}=\sum_{j=1}^\infty
j\,\bar{t}_j\,\mL^{j-1}-\sum_{j=1}^\infty
j\,\Big((\bar{t}_j+\bar{c}_j)-(t_j+c_j)\Big)\,(\mL^{j-1})_+.
\end{align}
In order to satisfy  \eqref{sati3} and \eqref{can} we introduce auxiliary functions of the form
\begin{align}\label{sati1}
\nonumber
\Psi&=\exp{\dfrac{1}{\epsilon}
\Big(\sum_{j=1}^\infty
t_j z^{j}+x\,\log{z}-\sum_{j\geq 1}\dfrac{1}{j\,z^j}S_{j+1}\Big)},\\\\
\nonumber
\overline{\Psi}&=\exp{\dfrac{1}{\epsilon}
\Big(\sum_{j=1}^\infty \bar{t}_j
z^{j}-(x+\epsilon)\,\log{z}-\bar{S}_0-\sum_{j\geq
1}\dfrac{1}{j\,z^j}\bar{S}_{j+1}\Big)},
\end{align}
and impose
\begin{align}\label{4.1.1}
\nonumber
\mL\,\Psi=z\,\Psi,\quad \mM\,\Psi=\epsilon\,\dfrac{\partial \Psi}
{\partial z},\\\\
\nonumber \bar{\mL}\,\overline{\Psi}=z\,\overline{\Psi},\quad
\overline{\mM}\,\overline{\Psi}=\epsilon\,\dfrac{\partial
\overline{\Psi}} {\partial z}.
\end{align}
Our aim is to determine $u,v,\mM$ and $\overline{\mM}$ from \eqref{4.1.1}. Now, with the help of \eqref{4.2}, we have that the equations
\eqref{4.1.1} for the Orlov operators read
\begin{align}\label{4.4}
\nonumber &\dfrac{x}{z}+\sum_{j\geq
2}\dfrac{1}{z^j}S_j=\sum_{j=1}^\infty
j\,\Big((\bar{t}_j+\bar{c}_j)-(t_j+c_j)\Big)\Big(\dfrac{1}{2z}\,T_{j[-1]}+\mathcal{O}\Big(\dfrac{1}{z^2}\Big)\Big),\\\\
\nonumber &-\dfrac{x+\epsilon}{z}+\sum_{j\geq
2}\dfrac{1}{z^j}\bar{S}_j=-\sum_{j=1}^\infty
j\,\Big((\bar{t}_j+\bar{c}_j)-(t_j+c_j)\Big)\Big(R_{j-1}+\dfrac{1}{2z}\,T_{j}+\mathcal{O}\Big(\dfrac{1}{z^2}\Big)\Big).
\end{align}
Matching the coefficients of $z^{-1}$ in both sides of these two
equations provides the same relation. Another relation is supplied
by identifying the coefficients of the constant terms in the
second equation of \eqref{4.4}. Hence we get a system of two
equations to determine $(u,v)$
\begin{equation}\label{4.5}\begin{cases}
\sum_{j=1}^\infty j\,\Big((\bar{t}_j+\bar{c}_j)-(t_j+c_j)\Big)
R_{j-1}=0,\\\\
\dfrac{1}{2}\,\sum_{j=1}^\infty
j\,\Big((\bar{t}_j+\bar{c}_j)-(t_j+c_j)\Big)\,T_{j[-1]}=x.
\end{cases}
\end{equation}
By equating the coefficients of the remaining powers of $z$ in
\eqref{4.4} we characterize the functions $S_j$ and $\overline{S}_j$ for
$j\geq 1$ in terms of $(u,v)$. Moreover, as it is proved below, the solution $(u,v)$ provided by \eqref{4.5} is of the
form
\[
u=\sum_{k\geq 0}\epsilon^k\,u^{(k)}(x,\ct,\dt),\quad v=\sum_{k\geq
0}\epsilon^{k}\,v^{(k)}(x,\ct,\dt),
\]
with $v^{(2k+1)}=0,\, \forall k\geq0$. Thus, by solving
\eqref{4.5} we  characterize operators $(\mL,\mM)$ and
$(\bar{\mL},\overline{\mM})$ which satisfy \eqref{tridiagonal} and
 all the requirements of Theorem 1. Therefore, they are
solutions of the Lax equations for the dispersionful 2-Toda
hierarchy.

We observe that, as it is noticed by Takasaki and Takebe in \cite{tt4}, solving the system of string equations
\eqref{tridiagonal} does not determine the coefficient $\bar{S}_0$ in \eqref{sati1} and therefore it does not determine a wave function $\overline{\Psi}$ of the dispersionful 1-Toda hierarchy.

\subsection{An iterative scheme for determining $(u,v)$}
It is convenient to write \eqref{4.5} in the form
\begin{equation}\label{hoin22}
\oint_{\gamma}\dfrac{d z}{2\pi i\,z}\, U_z\,\R(z)\, =0,\quad
\oint_{\gamma}\dfrac{d z}{2\pi i}\,U_z\,\T(z)\,
=-2\,(x+\epsilon),
\end{equation}
where $U$ denotes the function
\begin{equation}\label{U}
U(z,\ct,\dt):=\sum_{j=1}^\infty
\Big((t_j+c_j)-(\bar{t}_j+\bar{c}_j)\Big)\,z^j,
\end{equation}
and $\gamma$ is a large positively oriented closed path. Now by
using the first identity of \eqref{tri7} and the two equations of
\eqref{hoin22}  we find
\[
\oint_{\gamma}\dfrac{d z}{2\pi i}\,
U_z\,(\T+\T_{[-1]})=\oint_{\gamma}\dfrac{d z}{\pi i\,z}\,
(z-u)\,U_z\,\R= \oint_{\gamma}\dfrac{d z}{\pi i}\,
U_z\,\R=-4\,x-2\,\epsilon,
\]
so that \eqref{4.5} reduces  to a pair of equations involving
$\R$ only
\begin{equation}\everymath{\displaystyle}\label{hoin3}
\begin{cases}
\oint_{\gamma}\dfrac{d z}{2\pi i\,z}\,
U_z(z)\,\R(z)\, =0,\\\\
\oint_{\gamma}\dfrac{d z}{2\pi i}\,U_z(z)\,\R(z)\,
=-2\,x-\epsilon.
\end{cases}
\end{equation}
These equations together with the system \eqref{tri7}
\begin{equation}\label{RT}
\begin{cases}
\T_{[1]}+\T+\dfrac{2}{z}(u_{[1]}-z)\,\R_{[1]}=0,\\\\
\T^2-\dfrac{4}{z^2}\,v_{[1]}\,\R\,\R_{[1]}=1,
\end{cases}
\end{equation}
give rise an iterative scheme for characterizing $(u,v)$ as Taylor
series in $\epsilon$
\[
u=\sum_{k\geq 0}\epsilon^k\,u^{(k)}(x,\ct,\dt),\quad v=\sum_{k\geq
0}\epsilon^{k}\,v^{(k)}(x,\ct,\dt).
\]
The first step of the method is to determine
the expansions
\begin{equation}\label{sem}
\R(z)=\sum_{k\geq 0}\epsilon^k\,R^{(k)},\quad \T(z)=\sum_{k\geq
0}\epsilon^k\,T^{(k)},
\end{equation}
in terms of  $(u,v)$. It can be done by
equating the coefficients of  powers of $\epsilon$ in \eqref{RT}.
Indeed, the  coefficients of  $\epsilon^0$ leads to
\begin{equation}\label{RT0}
R^{(0)}\,=\,\frac{z}{\left((z-u^{(0)})^2-4v^{(0)}\right)^{\frac{1}{2}}},\quad
T^{(0)}\,=\,\frac{z-u^{(0)}}{\left((z-u^{(0)})^2-4v^{(0)}\right)^{\frac{1}{2}}},\end{equation}
and  the coefficients of $\epsilon^l$
($l\geq1$) yield the following system
\begin{align}\label{RTl}\everymath{\displaystyle}
\nonumber T^{(l)}-(z-u^{(0)})\frac{R^{(l)}}{z}&=\frac{1}{2}
\sum_{\scriptsize{\begin{array}{c}
i+j=l\\j\geq1\end{array}}}\Big(\frac{(-1)^j}{j!}\partial_x^jT^{(i)}+2u^{(j)}\frac{R^{(i)}}{z}\Big),\\
\\
\nonumber T^{(0)}T^{(l)}-4v^{(0)}\frac{R^{(0)}}{z}
\frac{R^{(l)}}{z}&=
2\sum_{\scriptsize\begin{array}{c}i+j+k=l\\j<l\end{array}}
\Big(\sum_{i_1+i_2=i}\frac{1}{i_2!}\partial_x^{i_2}v^{(i_1)}\Big)
\frac{R^{(j)}}{z}
\Big(\sum_{\scriptsize\begin{array}{c}k_1+k_2=k\\k_1<l\end{array}}
\frac{1}{k_2!}\partial_x^{k_2}\frac{R^{(k_1)}}{z}\Big)\\
\nonumber  &
-\frac{1}{2}\sum_{\scriptsize\begin{array}{c}i+j=l\\i,j\geq1\end{array}}T^{(i)}T^{(j)}.
\end{align}

Some comments concerning these formulas are in order
\begin{description}
\item[i)] The equations \eqref{RTl} determine each pair
$(T^{(l)},\,R^{(l)}/z)$ from $(T^{(j)},\,
R^{(j)}/z)$ with $j=0,1,\dots,l-1$.
\item[ii)] The equations
\eqref{RTl} are linear with respect to $T^{(l)}$, $R^{(l)}/z$.
Moreover, by taking into account \eqref{RT0}, we see that the
determinant of the coefficients of $T^{(l)}$ and $R^{(l)}/z$ in
\eqref{RTl} is
$$[(z-u^{(0)})^2-4v^{(0)}]^{\frac{1}{2}}.$$
\end{description}
Hence it follows that the functions $R^{(l)}/z$ can be written
as linear combinations of
$$\frac{z}{\left((z-u^{(0)})^2-4v^{(0)}\right)^{r+\frac{1}{2}}},\qquad
\frac{1}{\left((z-u^{(0)})^2-4v^{(0)}\right)^{r+\frac{1}{2}}},\qquad
r=1,2,\dots,l+1$$ with coefficients depending on $u^{(j)}$ and
$v^{(j)}$ with $j=0,1,\dots,l$ and their $x$-derivatives only.

Now let us go back to the system \eqref{hoin3} and find $(u,v)$. By substituting the $\epsilon$ expansion of $\R$ in
\eqref{hoin3} we get a system of two equations for each $R^{(l)}$
\vspace{0.3cm}
\begin{equation}\everymath{\displaystyle}\label{hoin4}
\begin{cases}
\oint_{\gamma}\dfrac{d z}{2\pi i\,z}\,
U_z(z)\,R^{(l)}(z)\, =0,\\\\
\oint_{\gamma}\dfrac{d z}{2\pi i}\,U_z(z)\,R^{(l)}(z)\,
=-2\,x\,\delta_{l0}-\delta_{l1},
\end{cases}
\end{equation}
which determine each pair $(u^{(l)},v^{(l)})$ recursively.
Furthermore, we can eliminate  the explicit dependence on
$(x,\ct,\dt)$ in the corresponding expressions since, by
differentiating with respect to $x$ the equations \eqref{hoin4}
for $l=0$
\begin{equation}\everymath{\displaystyle}\label{int0}
\begin{cases}
\frac{1}{2\pi
i}\oint_{\gamma}dz\frac{U_z}{\left((z-u^{(0)})^2-4v^{[0]}\right)^{\frac{1}{2}}}=0,\\\\
\frac{1}{2\pi
i}\oint_{\gamma}dz\frac{z\,U_z}{\left((z-u^{(0)})^2-4v^{(0)}\right)^{\frac{1}{2}}}=-2x,
\end{cases}
\end{equation}
all the integrals of the form
\begin{equation}\label{int}
\frac{1}{2\pi
i}\oint_{\gamma}dz\frac{U_z}{\left((z-u^{(0)})^2-4v^{(0)}\right)^{r+\frac{1}{2}}},\quad
\frac{1}{2\pi
i}\oint_{\gamma}dz\frac{z\,U_z}{\left((z-u^{(0)})^2-4v^{(0)}\right)^{r+\frac{1}{2}}},
\end{equation}
can be expressed in terms of $(u^{(0)},v^{(0)})$ and their
$x$-derivatives. We observe that \eqref{int} are the variables
introduced in \cite{amb1} to  determine the large $N$-expansion of
the hermitian matrix model.

Some important relations among the coefficients of the
semiclassical expansions under consideration are found by
realizing that given a solution $(u(\epsilon,x),
\,v(\epsilon,x),\,\R(\epsilon,z,x),\,\T(\epsilon,z,x))$ of
\eqref{hoin3}-\eqref{RT}, then
$$\everymath{\displaystyle}\begin{array}{lll}
\tilde{u}(\epsilon,x):=u(-\epsilon,x+\epsilon),& &\tilde{v}(\epsilon,x):=v(-\epsilon,x),\\  \\
\tilde{\R}(\epsilon,z,x):=\R(-\epsilon,z,x+\epsilon)&
&\tilde{\T}(\epsilon,z,x):=\T(-\epsilon,z,x+2\epsilon),
\end{array}$$
 satisfies \eqref{hoin3}-\eqref{RT} as well.
Thus, since the solution of \eqref{hoin3}-\eqref{RT} is uniquely
determined by $(u^{(0)},\,v^{(0)})$ we deduce that
$$\everymath{\displaystyle}\begin{array}{lll}
\tilde{u}(\epsilon,x)\,=\,u(\epsilon,x),&  &
\tilde{v}(\epsilon,x)\,=\,v(\epsilon,x),\\  \\
\tilde{\R}(\epsilon,z,x)\,=\,\R(\epsilon,z,x),&   &
\tilde{\T}(\epsilon,z,x)\,=\,\T(\epsilon,z,x).
\end{array}$$
Hence we find
\begin{equation}\label{const}\everymath{\displaystyle}\begin{array}{lllllll}
 u^{(2j-1)}&=&\frac{1}{2}\sum_{k=1}^{2j-1}\frac{(-1)^{k+1}}{k!}\partial_x^ku^{(2j-1-k)},&  &v^{(2j-1)}&=&0,\\  \\
R^{(2j-1)}&=&\frac{1}{2}\sum_{k=1}^{2j-1}\frac{(-1)^{k+1}}{k!}\partial_x^kR^{(2j-1-k)},&
&
T^{(2j-1)}&=&\frac{1}{2}\sum_{k=1}^{2j-1}\frac{(-1)^{k+1}2^k}{k!}\partial_x^k
T^{(2j-1-k)},
\end{array}\end{equation}
for $j=1,2,\dots$.

\subsubsection{Examples of calculations}
 Using \eqref{const} for $j=1$, it is immediately found that
\begin{equation}\label{RT1}
R^{(1)}\,=\,\frac{z\,\left( \left( z - u^{(0)} \right) \,u^{(0)}_x + 2\,v^{(0)}_x \right)
      }{2\,{\left( (z-u^{(0)})^2 - 4\,v^{(0)} \right)
        }^{\frac{3}{2}}},\quad
T^{(1)}\,=\,
\frac{4\,v^{(0)}\,u^{(0)}_x +
    2\,\left( z - u^{(0)} \right) \,v^{(0)}_x}{{\left( (z-u^{(0)})^2 - 4\,v^{(0)}  \right) }^{\frac{3}{2}}},
\end{equation}
and
\begin{equation}\label{sol1}
u^{(1)}=\frac{1}{2}u^{(0)}_x,\qquad v^{(1)}=0.
\end{equation}
With the help of \emph{Mathematica}, one
obtains
\begin{align}\label{R2}
\nonumber
\frac{R^{(2)}}{z}&=\dfrac{4(z-u^{(0)})u^{(2)}+8v^{(2)}+{u_x^{(0)}}^2+2v_{xx}^{(0)}}
{4\left((z-u^{(0)})^2-4v^{(0)}\right)^{\frac{3}{2}}}\\
\nonumber
\\
&+ \dfrac{\frac{7}{2}v^{(0)}{u_x^{(0)}}^2+
\frac{5}{2}(z-u^{(0)})v_x^{(0)}u_x^{(0)}+(z-u^{(0)})v^{(0)}u_{xx}^{(0)}+3{v_x^{(0)}}^2
+2v^{(0)}v_{xx}^{(0)}}
{\left((z-u^{(0)})^2-4v^{(0)}\right)^{\frac{5}{2}}}
\\
\nonumber
\\
\nonumber &+
\dfrac{10(z-u^{(0)})u_x^{(0)}v^{(0)}v_x^{(0)}+10{v^{(0)}}^2{u_x^{(0)}}^2+10v^{(0)}{v_x^{0}}^2}
{\left((z-u^{(0)})^2-4v^{(0)}\right)^{\frac{7}{2}}},
\end{align}
which leads to

$$\everymath{\displaystyle}\begin{array}{lll}
u^{(2)}&=&\frac{u_{xx}^{(0)}}{4} +
         \frac{v^{(0)}(7{u_x^{(0)}}^2u_{xx}^{(0)} -4u_{xx}^{(0)}v_{xx}^{(0)} - 2u_x^{(0)}\,v_{xxx}^{(0)}) +
    v_x^{(0)}( {u_x^{(0)}}^3 - 2u_x^{(0)}\,v_{xx}^{(0)} +
       2v^{(0)}u_{xxx}^{(0)}) }{24( {v_x^{(0)}}^2 - v^{(0)}{u_x^{(0)}}^2) }\\  \\
    & + &\frac{v^{(0)}u_x^{(0)}( {u_x^{(0)}}^4v_x^{(0)} +
      4v^{(0)}u_{xx}^{(0)}({u_x^{(0)}}^3  -2u_x^{(0)}v_{xx}^{(0)}) +
      4v_x^{(0)}( v^{(0)}{u_{xx}^{(0)}}^2 + {v_{xx}^{(0)}}^2 - {u_x^{(0)}}^2v_{xx}^{(0)}) ) }
         {24 {(  {v_x^{(0)}}^2-v^{(0)}\,{u_x^{(0)}}^2 )}^2} \\  \\
v^{(2)}&=&-\frac{ {v^{(0)}}^2u_x^{(0)}( {u_x^{(0)}}^5 +
4{u_x^{(0)}}^2v_x^{(0)}u_{xx}^{(0)} - 4v_{xx}^{(0)}({u_x^{(0)}}^3+
        2v_x^{(0)}u_{xx}^{(0)}) + 4u_x^{(0)}( v^{(0)}{u_{xx}^{(0)}}^2 +
           {v_{xx}^{(0)}}^2 ))   }{24{( {v_x^{(0)}}^2-v^{(0)}{u_x^{(0)}}^2 )}^2}  \\  \\
        & -&\frac{v^{(0)}( {u_x^{(0)}}^4 -
      3{u_x^{(0)}}^2v_{xx}^{(0)} +
      2u_x^{(0)}( 2\,v_x^{(0)}\,u_{xx}^{(0)} +
         v^{(0)}\,u_{xxx}^{(0)})  +
      2( v^{(0)}\,{u_{xx}^{(0)}}^2 + {v_{xx}^{(0)}}^2 -
         v_x^{(0)}\,v_{xxx}^{(0)} ) ) }{24({v_x^{(0)}}^2 -v^{(0)}{u_x^{(0)}}^2 )}
\end{array}$$
A further coefficient
can be easily computed by taking $j=2$ in \eqref{const}. Thus  we
obtain
$$u^{(3)}\,=\,\frac{1}{2}u_x^{(2)}-\frac{1}{24}u_{xxx}^{(0)},\quad v^{(3)}\,=\,0.$$

\subsection{The classical limit}

In the classical limit $\epsilon\rightarrow 0$ the functions $(u,v)$ reduce to the first terms $(u^{(0)},\,v^{(0)})$ of their semiclassical expansions and verify the equations of the dispersionless 1-Toda hierarchy
\begin{equation}\label{dto1}
\partial_{t_j} u=\dfrac{1}{2}\,\partial_x\,(r_{j+1}-u\,r_j),\quad
\partial_{t_j} v=v\,\partial_x\,r_{j},
\end{equation}
where $r_j$ are the coefficients of the Laurent expansion of
$R:=R^{(0)}$
\begin{equation}\label{d7b}
R :=\dfrac{z}{p-\bar{p}}=\dfrac{z}{\sqrt{(z-u)^2-4v}}=\sum_{k\geq
0}\dfrac{r_k(u,v)}{z^k},\quad r_0=1,
\end{equation}
and we have taken into account (see \eqref{RT0}) that $\T=T^{(0)}=(z-u)\,R/z$. Here $p:=p^{(0)}$ and $\bar{p}:=\bar{p}^{(0)}$ are given by
\begin{align}\label{d5}
\nonumber
&p(z)=\dfrac{1}{2}\Big((z-u)+\sqrt{(z-u)^2-4v}\Big)=z-u-\dfrac{v}{z}+\cdots\\\\
\nonumber
&\bar{p}(z)=\dfrac{1}{2}\Big((z-u)-\sqrt{(z-u)^2-4v}\Big)=\dfrac{v}{z}+\cdots.
\end{align}

According to \eqref{int0}, $(u,v)$ are determined by
\begin{equation}\label{hoin}\everymath{\displaystyle}
\begin{cases}
\oint_{\gamma}\dfrac{dz}{2\pi i}
\dfrac{U_{z}}{\sqrt{(z-u)^2-4v}}\, =0,\\\\
\oint_{\gamma}\dfrac{dz}{2\pi
i}\dfrac{z\,U_{z}}{\sqrt{(z-u)^2-4v}}\, =-2\,x,
\end{cases}
\end{equation}
which can be also expressed as \emph{hodograph} type equations
\begin{equation}\label{ho}\begin{cases}
\sum_{j=1}^\infty j\,\Big((\bar{t}_j+\bar{c}_j)-(t_j+c_j)\Big)
r_{j-1}=0,\\\\
\dfrac{1}{2}\,\sum_{j=1}^\infty
j\,\Big((\bar{t}_j+\bar{c}_j)-(t_j+c_j)\Big)\,r_j=x.
\end{cases}
\end{equation}
\subsection{The planar limit of the Hermitian matrix model}

From \eqref{ree3} the one-point correlator $W(z)$ is given by
\[
W(z)=\dfrac{1}{z}+\mathbb{M}_r(\epsilon,z,1,\ct)-\epsilon\,
\sum_{j\geq 1}\dfrac{1}{z^{j+1}}\dfrac{\partial
\mathbb{S}_r}{\partial t_j}(\epsilon,z,1,\ct),
\]
so that by using the first equations of \eqref{contwave1i} and
\eqref{4.3} one finds
\begin{align}\label{cor1}
\nonumber W(z)&=\sum_{j=0}^\infty
(j+1)\,\Big(\tilde{t}_{j+1}-\dfrac{\epsilon}{(j+1)\,z^{j+1}}\Big)\,
(\alpha_j+\beta_j\,p(z)-z^j)\\ \\
\nonumber &=-\sum_{j=1}^\infty
j\,\Big(\tilde{t}_{j}-\dfrac{\epsilon}{j\,z^{j}}\Big)\,
\Big(-\dfrac{1}{2}R_{j-1}+\dfrac{z}{2\,\R}\,(z^{j-2}\,\R)_\ominus-
\Big(\dfrac{z}{2\,\R}\,\T\,(z^{j-2}\,\R)_\oplus\Big)_\ominus\Big),
\end{align}
where
\[
\tilde{t}_j:=t_j+c_j.
\]
We are going  to show that the  solution of the dispersionless
1-Toda hierarchy determined by \eqref{ho} and \eqref{rest} describes the planar limit of the Hermitian matrix
model in the \emph{one-cut} case where the density of eigenvalues
\[
\rho(z)=M(z)\,\sqrt{(z-a)(z-b)},
\]
is supported on a single interval $[a,b]$. As it is known (see for instance \cite{mig}-\cite{eyn}) these
objects are related  to the
first term $W^{(0)}$ of the large $N$-expansion of $W$ in the form
\[
W^{(0)}=-\dfrac{1}{2}V_z(z)+i\pi\,\rho(z), \quad
V(z):=\sum_{k\geq 1}\tilde{t}_k\,z^k.
\]
According to
\eqref{RT0}, in the classical limit $\T=T^{(0)}=(z-u)\,R/z$ and then from
\eqref{cor1} it follows
\[
W^{(0)}=\dfrac{1}{2}\sum_{j=1}^\infty
j\,\tilde{t}_j\,r_{j-1}-\dfrac{1}{2}\sum_{j=1}^\infty
j\,\tilde{t}_j\,z^{j-1}+\dfrac{1}{2}(p-\bar{p})\sum_{j=2}^\infty
j\,\tilde{t}_j\,\Big(z^{j-2}\,R\Big)_\oplus,
\]
with $x=1$ in all $x$-dependent functions. Due to \eqref{rest}  the first hodograph
equation \eqref{ho} implies that the first term in the last equation
vanishes. Therefore the expressions for the density of eigenvalues
and the end-points of its support  provided the above solution of the
dispersionless 1-Toda hierarchy are
\begin{align}\label{den}
\nonumber \rho(z)&:=\dfrac{1}{2\pi i}\Big(\dfrac{V_z}{\sqrt{(z-a)(z-b)}}\Big)_\oplus\,\sqrt{(z-a)(z-b)},\\\\
\nonumber a&:=u-2\,\sqrt{v},\quad b:=u+2\,\sqrt{v},
\end{align}
where $x=1$ in all $x$-dependent functions. Moreover, from
\eqref{hoin}, they are determined by the equations
\begin{equation}\label{hoin2}\everymath{\displaystyle}
\begin{cases}
\oint_{\gamma}\dfrac{dz}{2\pi i}
\dfrac{V_{z}}{\sqrt{(z-a)(z-b)}}\, =0,\\\\
\oint_{\gamma}\dfrac{dz}{2\pi
i}\dfrac{z\,V_{z}}{\sqrt{(z-a)(z-b)}}\, =-2.
\end{cases}
\end{equation}
They coincide with the equations for the planar limit
contribution to
the partition function of the hermitian model
\cite{bipz}-\cite{eyn}.

\section{Critical points and the double scaling limit}

As we have seen the characterization of $(u,v)$ as  semiclassical expansions relies on the determination of smooth
leading terms $(u^{(0)},\,v^{(0)})$, which are defined implicitly by the hodograph equations \eqref{int0}. However, near critical points the functions $(u^{(0)},\,v^{(0)})$ are multivalued and have singular $x$-derivatives. Thus the semiclassical  expansions are not longer valid and a different procedure must be used. In this subsection
 we indicate how the so called \emph{double scaling limit} method (see for instance \cite{gin})
 can be formulated in our scheme.

 To simplify the discussion we set $u\equiv 0$ and
\begin{equation}\label{simm}
t_{2j-1}=c_j=0,\quad j\geq 1;\quad\bar{t}_j=\bar{c}_j=0,\quad j\geq
1,
\end{equation}
so that the Lax operator is of the form
\begin{equation}\label{trisim}
\mL=\Lambda+v\,\Lambda^{-1},
\end{equation}
and we are only considering the Toda flows associated with the even times $t_{2j}$. After eliminating   $\R$ in \eqref{tri7}, one sees that the
generating function $\U:=\T_{[-1]}$ satisfies the identity
\begin{equation}\label{tsim}
v\Big(\U+\U_{[-1]}\Big)\Big(\U+\U_{[1]}\Big)=z^2(\U^2-1).
\end{equation}
This leads to expansions of the form
\begin{equation}\label{usim}
\U=\sum_{j\geq 0}\dfrac{U_{2j}}{z^{2j}},\quad  \U=\sum_{k\geq
0}\epsilon^{2k}\,U^{(k)}.
\end{equation}
On the other hand, the system \eqref{4.5} reduces to
\begin{equation}\label{hosim}
-\,\sum_{j=1}^\infty j\,t_{2j}\,U_{2j}=
-\frac{1}{4\pi
i}\oint_{\gamma}dz\,V_z\,\U=x,
\end{equation}
where $\quad V=\sum_{k\geq 1}t_{2k}\,z^{2k}$.
Thus , the
solution $v$ is found from
 \eqref{tsim} and \eqref{hosim}.  In particular, the leading term $v^{(0)}$ is implicitly determined by
the hodograph equation
\begin{equation}\label{hh}
H(\bt_{even},v^{(0)})=x,
\end{equation}
where
\[
H(\bt_{even},v):=-\dfrac{1}{4\pi
i}\oint_{\gamma}dz\,V_z\,U^{(0)}=
-\frac{1}{4\pi
i}\oint_{\gamma}dz\frac{z\,V_z}{(z^2-4v)^{\frac{1}{2}}}.
\]

Given a general $m$-th order critical point $v_c:=v_c(\bt_{even})$ satisfying
\[
\dfrac{\partial H}{\partial v}\Big|_{v_c}=\ldots=\dfrac{\partial^{m-1}
H}{\partial v^{m-1}}\Big|_{v_c}=0,\quad \dfrac{\partial^{m} H}{\partial
v^{m}}\Big|_{v_c}\neq 0,
\]
the
method of the double scaling limit introduces a new small parameter
$\tilde{\epsilon}$ and a new variable $\tilde{x}$ given by
\begin{equation}\label{dl}
\tilde{\epsilon}:=\epsilon^{\frac{2}{2m+1}},\quad
x=H(v_c)+\tilde{\epsilon}^m\,\tilde{x},
\end{equation}
and generates solutions to \eqref{tsim} and \eqref{hosim} of the form
\begin{equation}\label{dl1}
 v=v_c\Big(1+\sum_{k\geq
1}\tilde{\epsilon}^{k}\,u^{(k)}\Big),\quad  \U=\sum_{k\geq
0}\tilde{\epsilon}^{k}\,\tilde{U}^{(k)}.
\end{equation}
To prove it, we first observe that $\epsilon\,\partial_x=\tilde{\epsilon}^{1/2}\,\partial_{\tilde{x}}$,
so that \eqref{tsim} can be rewritten as
\begin{equation}\label{tsim1}
v\,\sum_{n\geq 1}\,\tilde{\epsilon}^n\Big(\dfrac{4}{(2n)!}
\U\,\partial_x^{2n}\U+\sum_{k+l=2n
;\,k,l\geq 1} \dfrac{(-1)^k}{k!\,l!}\,\partial_x^{k}\U\,\partial_x^{l}\U \Big)=(z^2-4\,v)\,\U^2-z^2,
\end{equation}
and by substituting the expansions \eqref{dl1} in this identity and equating $\tilde{\epsilon}$-powers one can express each coefficient $\tilde{U}^{(n)}$ in the form
\begin{equation}\label{tesim2}
\tilde{U}^{(n)}=\sum_{r=1}^n\dfrac{z\,v_c^r\,G_{n,r}}{(z^2-4\,v_c)
^{\frac{2r+1}{2}}},
\end{equation}
where the coefficients $G_{n,r}$ are differential polynomials in
$u^{(k)},\,1\leq k\leq n-r+1$ and their $\tilde{x}$-derivatives . In
particular
\[
G_{n,1}= 2\,u^{(n)},
\]
and the first few $\tilde{U}^{(n)}$ are
$$\everymath{\displaystyle}\begin{array}{lll}
 \tilde{U}^{(0)}&=&
\frac{z}{(z^2-4v_c)^{\frac{1}{2}}}, \quad \quad \tilde{U}^{(1)}=
\frac{2\,v_c\,z\,u^{(1)}}{(z^2-4v_c)^{\frac{3}{2}}},\\  \\
\tilde{U}^{(2)}&=& \frac{2\,v_c\,z\,u^{(2)}}{{\left( z^2-4\,v_c \right)}^{\frac{3}{2}}} +
  \frac{2\,{v_c}^2\,z\,\left( 3\,{u^{(1)}}^2 + \partial_{\tilde{x}}^2u^{(1)} \right) }
   {{\left( z^2-4\,v_c  \right) }^{\frac{5}{2}}},\\  \\
\tilde{U}^{(3)}&=& \frac{2\,v_c\,z\,u^{(3)}}{{\left( z^2-4\,v_c
\right) }^{\frac{3}{2}}}  +
  \frac{{v_c}^2\,z\,\Big( 12\,u^{(1)}\,
        \left( 6\,u^{(2)} + \partial_{\tilde{x}}^2u^{(1)} \right)  + 12\,\partial_{\tilde{x}}^2u^{(2)} +
       \partial_{\tilde{x}}^4u^{(1)} \Big) }{6\,(z^2-4v_c)^{\frac{5}{2}}}\\  \\
       &  & +
  \frac{2\,{v_c}^3\,z\,\left( 10\,(u^{(1)})^3 + 5\,(\partial_{\tilde{x}}u^{(1)})^2 +
       10\,u^{(1)}\,\partial_{\tilde{x}}^2u^{(1)} + \partial_{\tilde{x}}^4u^{(1)} \right) }{{\left(z^2-4\,v_c  \right) }^{\frac{7}{2}}} .
\end{array}$$
Notice that $\tilde{U}^{(0)}(v)={U}^{(0)}(v)$.

By substituting \eqref{dl}-\eqref{dl1} in \eqref{hosim}  we get the system
\begin{equation}\label{sysdl}\everymath{\displaystyle}
\begin{cases}
\oint_{\gamma}dz\,V_z\,\tilde{U}^{(j)}=0,\quad j=1,\ldots,m-1
,\\\\
-\frac{1}{4\pi
i}\oint_{\gamma}dz\,V_z\,\tilde{U}^{(n)}=\delta_{nm}\,\tilde{x}
,\quad n\geq m.
\end{cases}
\end{equation}
Since $v_c$ is a $m$-th order critical point of \eqref{hh} we have
that
\[
\oint_{\gamma}dz\,\dfrac{z\,V_z}{(z^2-4\,v_c) ^{\frac{2j+1}{2}}}
=0,\quad j=1,\ldots,m-1.
\]
Hence, in view of \eqref{tesim2}, the first $m-1$ equations in
\eqref{sysdl}  are identically satisfied while the remaining ones
become
\begin{equation}\label{dd}
-\sum_{r=m}^n\frac{v_c^r\,G_{n,r}}{4\pi
i}\oint_{\gamma}dz\,\dfrac{z\,V_z}{(z^2-4\,v_c)
^{\frac{2r+1}{2}}}=\delta_{nm}\,\tilde{x} ,\quad n\geq m.
\end{equation}
For $n=m$ we get the equation which determines the  the leading
contribution $u^{(1)}$ in the double scaling limit
\begin{equation}\label{doug}
G_{m,m}(u^{(1)})=K_m\,\tilde{x}, \quad K_m^{-1}:={v_c}^m\,\oint
\frac{dz}{4\pi i}\frac{V_z z}{(z^2-4v_c)^{\frac{2m+1}{2}}}.
\end{equation}
For example
\begin{align*}
m=2,&\quad \quad 2\left( 3\,{u^{(1)}}^2 +
\partial_{\tilde{x}}^2u^{(1)}\right)=K_2\,\tilde{x};
\\ m=3,&\quad\quad
2\left( 10\,(u^{(1)})^3 + 5\,(\partial_{\tilde{x}}u^{(1)})^2 +
       10\,u^{(1)}\,\partial_{\tilde{x}}^2u^{(1)} + \partial_{\tilde{x}}^4u^{(1)}
       \right)=K_3\,\tilde{x}
\end{align*}
For $n\geq m+1$ the equations of the system \eqref{dd}  characterize
the coefficients $u^{(k)}$ for $k\geq 2$.

The differential equations \eqref{doug} for  $u^{(1)}$  are
essentially the stationary KdV equations \cite{doug}. Indeed, from
\eqref{tsim1} and taking into account \eqref{tesim2}  one gets
($G_i:=G_{i,i},\,G_i':=\partial_{\tilde{x}}G_i,...$)
\[
2v_c\sum_{i+j=m-1}G_i\,G_j''-\sum_{i+j=m}G_i\,G_j+4v_c u^{(1)}
\sum_{i+j=m-1}G_i\,G_j-v_c\sum_{i+j=m-1}G_i'\,G_j'=0,
\]
which, up to trivial rescalings, coincides with the equation
verified by  the coefficients of the expansion of the resolvent
diagonal $R$ of the Sch\~odinger operator
\[
R\,R''-2(z^2-u)\,R^2-\dfrac{1}{2}R'^2+2\,z^2=0,\quad R=1+\sum_{j\geq 1}\dfrac{R_j}{z^{2j}}.
\]

\vspace{0.3cm}

\noindent {\bf Acknowledgements}

\vspace{0.3cm} The authors  wish to thank the  Spanish Ministerio de
Educaci\'on y Ciencia (research project FIS2005-00319) for its
finantial support. This work is also part of the MISGAM programme of
the European Science Foundation.


\vspace{0.5cm}


\begin{thebibliography}{99}

\bibitem{rus} A. Gerasimov, A. Marshakov, A. Mironov, A. Morozov and
A. Orlov, Nuc. Phys. B {\bf 357},  565 (1991)
\bibitem{mar} E. J. Martinec, Comm. Math.
Phys. {\bf 138} , 437 (1991)

\bibitem{avm} M. Adler and P. Van Moerbeke, Comm. Math.
Phys. {\bf 203} , 185 (1999); Comm. Math. Phys. {\bf 207} , 589
(1999)
\bibitem{fok} A. S. Fokas, A. R. Its and V. Kitaev, Uspekhi Mat. Nauk
{\bf 45} , 135 (1990) (in Russian), translation in Russian Math. Surveys {\bf 45} , 155 (1990) ; Comm. Math. Phys. {\bf 147} , 395
(1992)

\bibitem{amb1} J. Ambj{\o}rn, L. Chekhov  and
Yu. Makeenko , Phys. Lett. B {\bf 282},  341 (1992)

\bibitem{amb} J. Ambj{\o}rn, L. Chekhov , C. F. Kristjansen and
Yu. Makeenko , Nuc. Phys. B {\bf 404},  127 (1993)

\bibitem{ble} P. Bleher and A. Its, Annals Math. {\bf 150} , 185
(1999)
\bibitem{dei} P. Deift, \emph{Orthogonal Polynomials and Random Matrices: A Riemann-Hilbert Approach},
 Courant Lecture Notes in Mathematics {\bf 3}, Amer. Math. Soc. Providence, RI, (1999)

\bibitem{ueta} K. Ueno and T. Takasaki, \textit{Toda lattice hierarchy} in \textit{Group representations and systems of differential
equations},  Adv. Stud. Pure Math. \textbf{4}, 1, North Holland,
Amsterdam (1984)
\bibitem{tt4}  K. Takasaki and T. Takebe,  Rev. Math.
Phys. {\bf 7},  743 (1995)
\bibitem{tak5} K. Takasaki, Comm. Math.
Phys. {\bf 170} , 101 (1995)

\bibitem{dub1} B. Dubrovin and Y. Zhang, \emph{Normal Forms of Integrable PDEs, Frobenius Manifolds and
Gromov-Witten invariants } math/0108160

Comm. Math. Phys. {\bf 203} , 185 (1999); Comm. Math. Phys. {\bf
250} 161 (2004)


\bibitem{dub2} G. Carlet, B. Dubrovin and Y. Zhang,
Moscow Math. J. {\bf 4} 313 (2004) 313-332.

\bibitem{dub3} B. Dubrovin and Y. Zhang, Comm. Math.
Phys. {\bf 203} , 185 (1999); Comm. Math. Phys. {\bf 250} 161 (2004)

\bibitem{mel3} L. Martinez Alonso and E. Medina, Phys. Lett. B {\bf 610}, 227 (2005)

\bibitem{mel4}  L. Mart\'{\i}nez Alonso  y  E. Medina, Phys. Letters B {\bf 641},
466 (2006)
\bibitem{mano1}  M. Ma\~{n}as, E. Medina and L. Martinez Alonso, J. Phys. A: Math. Gen. {\bf 39}, 2349 (2006)

\bibitem{mano2}  L. Martinez Alonso, M. Ma\~{n}as and  E. Medina, J. Math. Phys. {\bf 47}, 83512 (2006)


\bibitem{che} I. V. Cherednik, Funct. Anal. Appl. {\bf 12:3}, 45 (Russian), 195 (English)  (1978)

\bibitem{wil} G. Wilson, Quart. J. Math. Oxford {\bf 32}, 491 (1981)

\bibitem{kup} B. A. Kuperschmidt, \emph{Discrete Lax equations and Differential-Difference
Calculus}, Aster\'{i}sque {\bf 123} (1989)

\bibitem{zab1} P. W. Wiegmann and P. B. Zabrodin,  Comm. Math.
Phys. {\bf 213} , 523 (2000)

\bibitem{zab2} M. Mineev-Weinstein, P. Wiegmann and A. Zabrodin,
Phys. Rev. Lett. {\bf 84},  5106

\bibitem{zab3} A. Boyarsky, A. Marsahakov, O. Ruchaysky,
P. Wiegmann and A. Zabrodin, Phys. Lett.B {\bf 515},  483 (2001)

\bibitem{zab4} O. Agam, E. Bettelheim, P. Wiegmann and A. Zabrodin,
Phys. Rev. Lett. {\bf 88},  236801 (2002)

\bibitem{zab5} I. Krichever, M. Mineev-Weinstein, P. Wiegmann and A. Zabrodin,
Physica D {\bf 198},  1 (2004)

\bibitem{zab6} A. Zabrodin,
Teor. Mat. Fiz.  {\bf 142}, 197 (2005)


\bibitem{zab7} R. Teodorescu, E. Bettelheim, O. Agam, A. Zabrodin
and P. Wiegmann, Nuc. Phys. B {\bf 700},  521 (2004); Nuc. Phys. B
{\bf 704},  407 (2005)


\bibitem{kaz} V. Kazakov and A. Marsahakov,
J. Phys. A {\bf 36}, 3107  (2003)


\bibitem{bipz} E. Br\'ezin, C. Itzikson, G. Parisi and B. Zuber, Comm. Math. Phys. {\bf 59} , 35 (1978)

\bibitem{biz} D. Bessis, C. Itzikson, G. Parisi and B. Zuber, Adv. in Appl. Math.  {\bf 1} , 109 (1980)


\bibitem{iz} C. Itzikson and B. Zuber, J. Math. Phys. {\bf 21} , 411 (1980)

\bibitem{mig} A. A. Migdal, Phys. Rep. {\bf 102}, 199 (1983)

\bibitem{eyn} B. Eynard, \emph{An Introduction to Random Matrices}, lectures given at Saclay, October 2000, http://www-spht.cea.fr/articles/t01/014/.

\bibitem{gin} P. Di Francesco, P. Ginsparg and Z. Zinn-Justin ,\emph{ 2D Gravity and Random Matrices} hep-th/9306153.

\bibitem{doug} M. Douglas, Phys. Lett.B {\bf 238},  176 (1990)



\end{thebibliography}
\end{document}